\documentclass{optica-article}

\journal{opticajournal} 

\articletype{Research Article}

\usepackage{layouts}
\usepackage{amsmath,bm}
\usepackage{todonotes}

\let\vec\bm

\begin{document}


\title{When can we see micromotion? Experimental and theoretical analysis of the ChiSCAT scheme}


\author{Andrii Trelin,\authormark{1,2} Jette Abel,\authormark{3} Christian Rimmbach,\authormark{2,4} Robert David,\authormark{2,4} Andreas Hermann,\authormark{3,5} Friedemann Reinhard,\authormark{1,2,*}}

\address{\authormark{1}Institute of Physics, University of Rostock, 18059 Rostock, Germany\\
\authormark{2}Department of Life, Light, and Matter of the Interdisciplinary Faculty at Rostock University, 18059 Rostock, Germany\\
\authormark{3}Translational Neurodegeneration Section „Albrecht Kossel“, Department of Neurology, University Medical Center Rostock, University of Rostock, Rostock, Germany\\
\authormark{4}Department of Cardiac Surgery, Rostock University Medical Centre, 18057 Rostock, Germany\\
\authormark{5}Center for Transdisciplinary Neurosciences Rostock (CTNR), University Medical Center Rostock, Rostock, Germany

}

\email{\authormark{*}friedemann.reinhard@uni-rostock.de}

\begin{abstract*} 
We present an in-depth analysis of ChiSCAT, a recently introduced interferometric microscopy scheme to detect recurring micromotion events in cells. Experimentally, we demonstrate that illumination with low-coherence sources can greatly improve the robustness of the scheme to vibrations. Theoretically, we analyze the performance of ChiSCAT under various noise models, in particular photon shot noise and noise dominated by cellular motions other than the signal. We finally propose ways to improve performance, especially in a setting dominated by cell motions, and conclude with an outlook on potential future directions.
\end{abstract*}

\section{Introduction}


Some evidence suggests that action potentials of electro-active cells like neurons or cardiomyocytes can be detected by label-free interferometric microscopy, because they are accompanied by "intrinsic optical signals", such as a nanometer-scale motion of the cell membrane \cite{hill49,cohen68,iwasa80,laporta1990recording,stepnoski1991noninvasive,macvicar1991imaging,kim2007mechanical,laporta2012interferometric,badreddine2016real,yang2018imaging,ling2018full,ling2020high,mueller2014quantitative,mosbacher1998voltage,gonzalez2016solitary}. 

Recently, we proposed ChiSCAT \cite{trelin2024chiscat}, a combination of interferometric microscopy scheme with eigendecomposition-based algorithm, which allowed experimental detection of APs in a non-neuronal cell culture of blebbistatin-paralyzed cardiomyocytes.  

In this article we aim to analyze the technical and fundamental limits of this scheme, with insights from the latter that generalize to a wide range of other interferometric microscopy schemes. 

Our work is organized as follows: we firstly analyze the experimental noise in our setup, finding that a major improvement is possible by using low-coherence light for the illumination. We go on to theoretically analyze the sensitivity that ChiSCAT can reach. Since this depends critically on the type of noise the scheme is subjected to, we explicitly compare two prototypical settings: photonic shot noise and noise from cellular background motions other than the signal. We find that cell noise presents a more serious challenge than shot noise, and propose ways to improve the robustness of the ChiSCAT scheme in this setting.

\begin{figure}[htbp]
\centering\includegraphics[width=0.75\textwidth]{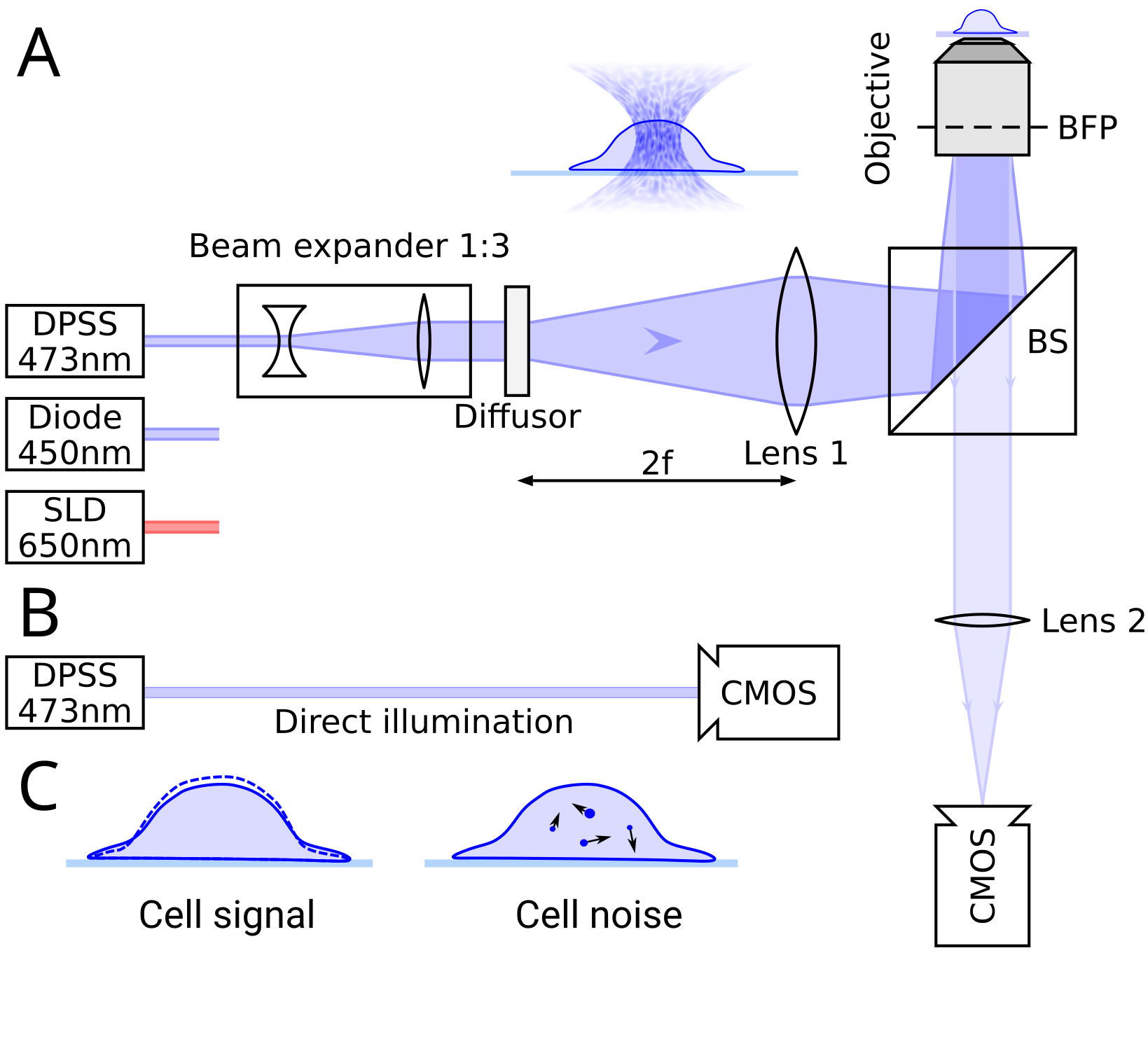}
\caption{(a) Interferometric cell imaging setup (ChiSCAT). The beam from one of the sources (DPSS laser at 473 nm, diode laser at 450 nm, or SLD at 650 nm) is expanded and passed through a holographic diffuser to create a random light field in the cell plane. The speckle pattern reflected from the cell and the coverslide interferes to generate the video recorded with a high-speed CMOS camera. (b) To measure inherent laser and camera noise laser was set to illuminate camera directly (Direct illumination). (c) The ultimate goal of this paper is to find criteria at which the cell signal can be reliable disentangled from the cell noise.}
\label{fig:setup}<
\end{figure}

\section{Experimental characterization of noise sources}

The task of detecting action potentials (APs) in electrically active cells from an interferometric microscopy recording is complicated by various noise sources. Shot noise, originating from the inherent statistics of photon detection, defines a fundamental limit of motion sensitivity for any interferometric scheme. Even if detection is possible within the shot noise limit, other noise sources can dominate and mask an AP signal. For example, cell motion noise, which arises from the inherent random micromotion of the cellular structures can mask a sub-nanometer AP signal, and so can vibrations of the setup. 
The ChiSCAT setup (Fig. \ref{fig:setup}), used in our experiments is based on wide-field interferometric scattering microscopy (iSCAT \cite{kukura2009high}) and uses coherent light to illuminate the cell. The light wavefront is randomized by a holographic diffuser to create a 3D random light field in the cell volume and to employ the full numerical aperture (NA) of the microscope for illumination. The observed signal is recorded by the high-speed camera (KronTech Chronos 2.1-HD) and consists of a superposition of speckle patterns arising from different components of the cell, interfering with light reflected at the cover slide interface, making it interferometrically sensitive to cellular motion. 

Thus, it is critical to develop an understanding of the different noise components and their properties. To obtain such an understanding, we will now compare ChiSCAT measurements with and without cells, in order to gauge cell noise, and using different illumination sources, in order to gauge their intensity fluctuations and their sensitivity to vibrations. The impact of vibrations is expected to depend on the illumination source. A source with a short coherence length will only create interference between closely spaced surfaces, i.e. the cell and the cover slide. In contrast, a source with a long coherence length interference can occur even between widely separated vibrating surfaces, such as different optical elements in the beam path. 

\begin{figure}[htbp]
\centering\includegraphics[width=\textwidth]{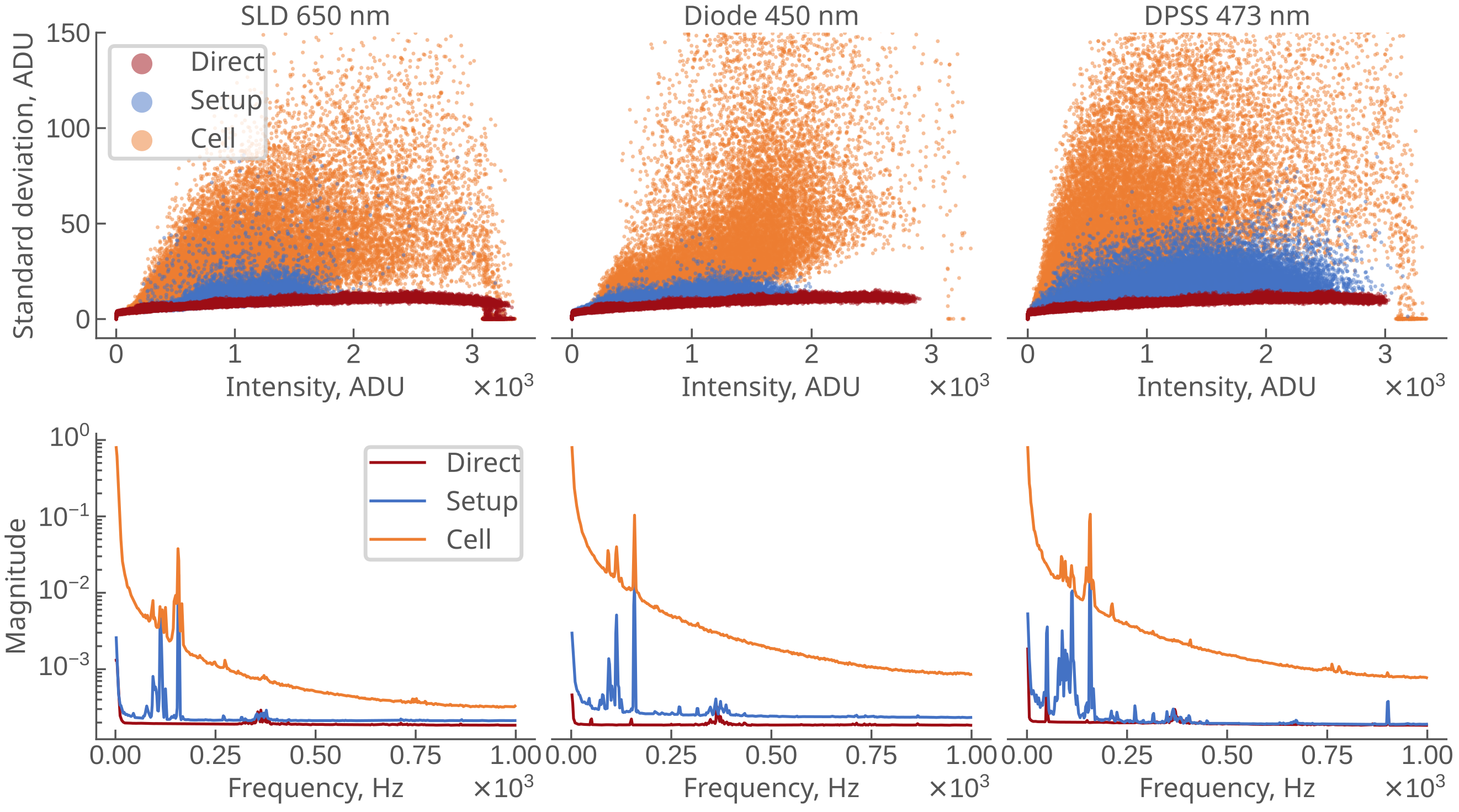}
\caption{Results of laser and cell noise characterization. Top row: dependence of the pixel standard deviation on intensity for every pixel of the camera sensor for different kinds of illumination and different versions of experiment. Bottom row: average Fourier transformed time traces from every pixel for different versions of the experiment and kind of illumination. ADU: analog-to-digital units.}
\label{fig:measurement_result}
\end{figure}

Specifically, the ChiSCAT setup  (Fig. \ref{fig:setup}) was utilized to perform measurements with different illumination sources, namely a diode-pumped 473 nm laser (DPSS, CNI Lasers MBL-FN-473-50mW, coherence length $\approx$25 mm), a 450 nm laser diode (Thorlabs LP450-SF25, coherence length $\approx$50 $\mu$m), and a 650 nm superluminescent diode (SLD, Thorlabs SLD650T, coherence length $\approx$25 $\mu$m). To manage vibrations, the whole setup was built on top of high-quality optical table (Thorlabs T1020Q) and every equipment that might generate vibrations was mechanically decoupled. Three measurements were conducted for each type of laser: directly illuminating the camera sensor (direct),  ChiSCAT microscopy without the cell but with a Petri dish containing water (setup), and with a living neuron that was not electrically active (cell). Induced pluripotent stem cells (iPSC) derived human neurons have been used in this experiment. All measurements were performed by continuously running the lasers at full power, allowing a 10-minute warm-up period before measurement to minimize noise. Appropriate illumination intensity was set using neutral density and polarization filters. Results of these experiments are presented in Fig. \ref{fig:measurement_result}.

In the upper row, the temporal standard deviation of every pixel of the high-speed recording is plotted against the corresponding pixel intensity. It can be observed that direct illumination of the camera sensor results in an almost perfect narrow band, demonstrating a $\sqrt{I}$ behavior, as predicted by the properties of shot noise. When the laser passes through the complete microscopy setup without a cell, the standard deviation increases. This is attributed to the sensitivity of the interferometric setup to vibrations, which is evident in the Fourier-transformed time traces (bottom row) as narrow peaks corresponding to different vibrational modes of the setup. The increase in noise for the blue DPSS laser illumination is the most pronounced, due to the long coherence length of that laser, which captures interference between reflections from more distant surfaces, such as between lenses in the objective.

However, an even more significant difference occurs when a living cell is introduced into the setup. The scatter plots demonstrate an increase in the pixel standard deviation by an order of magnitude. Additionally, the spectra indicate the presence of a Brownian noise-like background, reflecting the actual Brownian motion occurring within the cell. This leads to the conclusion that while the setup operates close to the shot noise limit, the noise introduced by the cell is considerably higher than this limit.

\section{Theoretical limits for AP detection}

\subsection{Shot noise limited case}
To understand the fundamental limitations of ChiSCAT approach, it is necessary to derive theoretical criteria of signal recovery. We will first consider the simplest case, that the AP detection performance of ChiSCAT is limited by shot noise. In this assumption, the AP signal is the only micromotion produced by the cell, and the only noise source present is the photon shot noise in the video recording.
In this subsection, we will develop a model to analyze under which conditions the AP signal can be recovered from the measurements in this setting.

\paragraph{Signal model and assumptions}

\begin{enumerate}
    \item In ChiSCAT the reflected laser intensity is recorded as a video $F$.
    By flattening the frame dimension, this video can be represented as matrix of size $(N_{\text{fr}}, N_{\text{px}})$, that consists of a static background $B$, an action potential pattern $P$ that occurs recurrently, and shot noise $N$:

    \[
    F = B + P + N
    \]

    To simplify analysis, it can be assumed that the video is centered by subtracting the constant background ($B=0$). 

    \item The action potential signal $P$ can be decomposed into a spatial pattern $\vec{p}$ of size $(N_{\text{px}})$ and a temporal pattern $\vec{a}$ of size $(N_{\text{fr}})$:

    \[
    P = \vec{a} \vec{p}^{\intercal}
    \]

    The video is supposed to be scaled such that the amplitude of $\vec{a}$ is 1. The spatial pattern is assumed to be binary, with $p_{i} \in \{-1, 1\}$, where both values have an equal probability of 0.5. This reflects the varying sensitivity of the pixels to cellular motion. The pattern is zero-centered as it does not affect the total intensity of the observed signal.

    \item Every element of the noise matrix $N_{ij}$ is independently and identically distributed (i.i.d.), approximately Gaussian, with a variance of $\sigma_{\text{1px}}^{2}$. Due to the above rescaling to unit signal amplitude, $1/\sigma_{\text{1px}}$ is the signal-to-noise ratio (SNR) obtained if the AP is reconstructed from a single-pixel time trace, whereas $\sigma_{\text{1px}}$ is the noise in this single-pixel trace.
\end{enumerate}

\paragraph{Algorithm}
The ChiSCAT algorithm, proposed in \cite{trelin2024chiscat}, relies on the eigendecomposition of the Gram or covariance matrix $G=FF^{\intercal}$ (both terms used interchangeably). Note that the original algorithm contains an additional step of erasing the diagonal of $G$, which we will neglect in the following derivation.
The algorithm makes an assumption that the first eigenvector $\vec{v}$ of $G$ approximates the true temporal pattern $\vec{a}$. To demonstrate this, the matrix $G$ can be decomposed into components as follows:

\begin{align}
G & =FF^{\intercal}=(P+N)(P+N)^{\intercal}\label{eq:signal_model} \\ 
  & =PP^{\intercal}+PN^{\intercal}+NP^{\intercal}+NN^{\intercal}.
\nonumber 
\end{align}

In the absence of the noise ($N=0$), it holds that

\[
G=PP^{\intercal}=(\vec{a}\vec{p}^{\intercal})(\vec{a}\vec{p}^{\intercal})^{\intercal}=\vec{a}\vec{p}^{\intercal}\vec{p}\vec{a}^{\intercal}=\|\vec{p}\|^{2}\vec{a}\vec{a}^{\intercal}
\]
i.e. that $G$ is essentially the outer product of $\vec{a}$ with itself, a rank-1 matrix for which the first eigenvector corresponds to $\vec{a}$  (up to a scaling constant). However, this ideal scenario does not hold when noise is present. Under such conditions, the first eigenvector might or might not correlate with the true signal $\vec{a}$. We will refer to the random matrix theory to
understand conditions at which largest eigenvector actually bears
information about true AP.

\paragraph{Recovery criterion}

\begin{figure}[htbp]
\centering\includegraphics[width=0.5\textwidth]{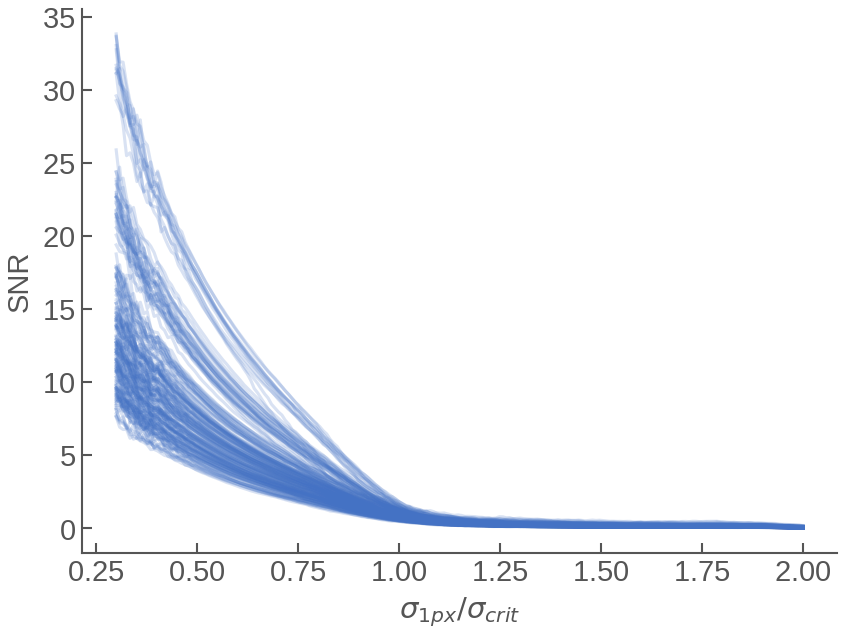}

\caption{Numerically simulated SNR of recovered signal at different values of $\sigma_{\text{1px}}/\sigma_{\text{crit}}$}
\label{fig:SNR-for-recovered}
\end{figure}

In scenarios where the observed matrix is the sum of a low-rank matrix and random noise, spiked random matrix theory \cite{perry2018optimality} can be utilized to establish criteria for signal recovery, specifically criteria for when the signal-to-noise ratio (SNR) of the recovered signal exceeds one. A useful model in this context is the spiked Wigner model, a sum of rank-1 outer product matrix $\vec{x} \vec{x}^{\intercal}$ with noise, given by:

\begin{equation}
Y = \lambda \vec{x} \vec{x}^{\intercal} + \frac{1}{\sqrt{n}} W \label{eq:wigner_model}
\end{equation}
where $\vec{x}$ is a unit vector, $W$ represents a symmetric random matrix of size $n\times n$ with independent and identically distributed (i.i.d) random entries with unit variance. This model closely resembles the expression in Eq. \ref{eq:signal_model}. Specifically, $\lambda \vec{x} \vec{x}^{\intercal}$ is comparable to $PP^{T}$, the noise matrix $W$ is analogous to $NN^{\intercal}$ 
,  while $n$ corresponds to $N_{\text{fr}}$. The cross components $PN^{\intercal} + NP^{\intercal}$ are sparse (since $\vec{a}$ is sparse) and small in the low SNR regime, which is our primary focus. We will thus neglect them, and will later confirm this assumption by a numerical simulation (Fig. \ref{fig:SNR-for-recovered}).


The most important finding from \cite{perry2018optimality} indicates that recovery is possible when $\lambda > 1$, which \textit{de-facto} answers the question. To explicitly connect Eq. \ref{eq:signal_model} to the spiked Wigner model one can multiply both sides of  by $\frac{1}{\sqrt{N_{\text{fr}}}}\frac{1}{\sqrt{N_{\text{px}}}\sigma_{\text{1px}}^{2}}$, leading to

\[
\frac{1}{\sqrt{N_{\text{fr}}}}\frac{1}{\sqrt{N_{\text{px}}}\sigma_{\text{1px}}^{2}}G=\frac{1}{\sqrt{N_{\text{fr}}}}\frac{\|\vec{p}\|^{2}\|\vec{a}\|^{2}}{\sqrt{N_{\text{px}}}\sigma_{\text{1px}}^{2}}\Big(\frac{\vec{a}}{\|\vec{a}\|}\frac{\vec{a}^{\intercal}}{\|\vec{a}\|}\Big)+\frac{1}{\sqrt{N_{\text{fr}}}}\frac{1}{\sqrt{N_{\text{px}}}\sigma_{\text{1px}}^{2}}NN^{\intercal}.
\]

Then, the matrix $\frac{1}{\sqrt{N_{\text{px}}}\sigma_{\text{1px}}^{2}}NN^{\intercal}$ is a random symmetric matrix with unit standard deviation (which follows from the fact that entries in $NN^{\intercal}$ have standard deviation $\sqrt{N_{\text{px}}}\sigma_{\text{1px}}^2$ since it's a sum of $N_{\text{px}}$ products of independent random variables in $N$, each having standard deviation $\sigma_{\text{1px}}$). Note that $\frac{\vec{a}}{\|a\|}$ is a unit vector, thus the rescaled equation can be identified with Eq. \ref{eq:wigner_model}. That translates the recovery criterion to:
\[
\lambda=\frac{1}{\sqrt{N_{\text{fr}}}}\frac{\|\vec{p}\|^{2}\|\vec{a}\|^{2}}{\sqrt{N_{\text{px}}}\sigma_{\text{1px}}^{2}}=\frac{\|\vec{p}\|^{2}\|\vec{a}\|^{2}}{\sqrt{N_{\text{px}}}\sigma_{\text{1px}}^{2}}>1
\]
which under the assumptions $\|\vec{p}\|^{2}=N_{\text{px}}$, $\|\vec{a}\|^{2}=N_{\text{ap}}$
can be reformulated as 
\begin{equation}    
\sigma_{\text{1px}}<\frac{\sqrt{N_{\text{ap}}}\sqrt[4]{N_{\text{px}}}}{\sqrt[4]{N_{\text{fr}}}}=:\sigma_{\text{crit}}.
\label{eq:max_sigma}
\end{equation}
which also matches the empirically observed one in \cite{trelin2024chiscat}.

\paragraph{Numerical simulation}
To further verify the validity of the derivation, a numerical simulation was performed to analyze dependence of the SNR of signal, recovered with ChiSCAT algorithm on $\sigma_\text{1px}$. Specifically, a vector $\vec{p}$ of length $N_{\text{px}}$ was randomly generated from a $\{-1,1\}$ distribution. A vector $\vec{a}$ was simulated by randomly setting $N_{\text{ap}}$ entries to 1 in an array composed of $N_{\text{fr}}$ zeros. The video $F$ was then created as the outer product of the spatial and temporal patterns $F=\vec{a}\otimes\vec{p}+N$, where each element of the matrix $N$ follows a normal distribution $\sim\mathcal{N}(0,\sigma_{\text{1px}})$. With the matrix $F$ obtained, the Gram matrix was computed as $G=FF^{\intercal}$, from which the first eigenvector $\vec{v}$ was determined. To calculate the numerical SNR, the eigenvector was decomposed into the linear basis of $\vec{a}$ and the remaining noise as $\vec{v}=\alpha\vec{a}+\vec{\varepsilon}$.
Consequently, the final simulation result, SNR, is expressed as $\text{SNR}=\alpha/\text{std}(\vec{\varepsilon})$. 

To assess the validity of Eq.~\ref{eq:max_sigma}, simulations were performed for randomly sampled parameters with $N_{\text{ap}}$ ranging from 3 to 200, $N_{\text{fr}}$ from 100 to 2000, and $N_{\text{px}}$ from 2000 to 5000. For each combination of $N_{\text{ap}}, N_{\text{px}}, N_{\text{fr}}$, the value $\sigma_{1px}$ was varied from $0.25\sigma_{\text{crit}}$ to $2\sigma_{\text{crit}}$. The results of this simulation are illustrated in Fig.~\ref{fig:SNR-for-recovered}, which clearly show a transition happening around $\sigma_{\text{1px}}=\sigma_{\text{crit}}$.

\paragraph{Implications}

For a realistic scenario one can assume $N_{\text{ap}}=10, N_{\text{fr}}=1000$ (given that the speckle signal is stable on the timescale on the order of 1 s and the camera framerate is on the order of 1000 FPS), $N_{\text{px}}=2000$,
where $N_{\text{px}}$ is defined by the number of resolved points in a 10 \(\mu m\) cell. In this case, $\sigma_{\text{crit}} \approx 3.7$. ChiSCAT could thus reveal an action potential even if its amplitude is $3.7$ times less than  the noise in a single-pixel time trace. 
For larger cells, such as Purkyně cells which range in size from 60 to 80 \(\mu m\) \cite{ghez1991cerebellum} and correspond to approximately 60,000 resolved points, signal detection is achievable when \(\sigma_{\text{1px}}<8.8\).

\subsection{Cell noise limited case}



Shot noise defines the ultimate fundamental limit of sensitivity, but might not be reached in practice, since cellular motions other than the AP (arising from the random movement or active transport of scatterers within the cell) can potentially mask the signal. To assess the effect of such cell noise, we will now study a modified signal model where shot noise in the detection is completely neglected, but multiple motion patterns are present in the cell, aiming to analyze recovery criteria in this case. 

\paragraph{Signal model and assumptions}

The model outlined previously considers only shot noise as the primary noise source. To take into account contributions from randomly moving scatterers, we model cell noise
similar to the representation of the action potential signal as the outer product of a spatial pattern vector and a temporal amplitude vector. Each particle contributes a unique random spatial pattern, $\vec{p}_{i}$, and a temporal amplitude, $\vec{a}_{i}$. It is assumed that the action potential signal corresponds to the first component $\vec{a}_{1}$, so that the components $i>2$ are arbitrary time series, not limited to sparse signals of discrete spikes. Consequently, the observed video $F$ is the sum of the contributions from all patterns:

\[
F = \sum_{i=1}^{N_{\text{c}}} \vec{a}_{i} \vec{p}_{i}^{\intercal} = AP
\]
where \(N_{\text{c}}\) represents the total number of signal components, matrix $A$ ($N_{\text{fr}} \times N_{\text{c}}$) contains temporal amplitudes in its columns, and matrix $P$ ($N_{\text{c}} \times N_{\text{px}}$) contains spatial patterns in its rows. The Gramian is then expressed as

\[
G = (AP)(AP)^{\intercal} = AP P^{\intercal} A^{\intercal}.
\]

Since each pattern is independent and zero-centered, assuming $N_{\text{px}} > N_{\text{c}}$, the following approximation holds:

\[
P P^{\intercal} \approx kI
\]
where the scaling constant $k = \mathbb{E}\|\vec{p}_{i}\|^{2} = N_{\text{px}}$ represents the expected magnitude of the patterns. Thus, the Gramian matrix approximates the covariance matrix of the amplitudes (up to a constant factor):

\begin{equation}
G \approx N_{\text{px}} A A^{\intercal}\label{eq:gramian_is_A}
\end{equation}

This implies that the outcome of the ChiSCAT algorithm is essentially a principal component analysis (PCA) of the matrix $A$. We will refer to the principal components as PC$i$ in the further text. It is safe to assume that motion of every particle in the cell is independent, which implies that the amplitude vectors $\vec{a}_i$ are approximately orthogonal. It is further assumed that the total variance of the noise is known (since it can be measured experimentally):

\[
\sigma_{\text{tot}}^{2} = \sum_{i=2}^{i=N_{\text{c}}} \sigma_{i}^{2}
\]
where $\sigma_{i}^{2}$ represents the variance of $\vec{a}_{i}$.

\paragraph{Recovery criterion}

Let us firstly focus on the case of $N_{\text{c}}=2$, i.e., to have just one noise component. In this case the problem becomes relatively straightforward. The matrix $A=[\vec{a}_{1},\vec{a}_{2}]$ is composed of the AP signal $\vec{a}_{1}$ and a noise vector $\vec{a}_{2}$. To faithfully recover the signal, the signal has to define the direction of the first eigenvector, i.e. the direction of the largest variance. The recovery criterion thus reads
$$\text{var}(\vec{a}_{1})>\text{var}(\vec{a}_{2})=\sigma_{\text{tot}}^{2}$$
$\text{var}(\vec{a}_{2})$, meant to be the temporal variance of the components of $\vec{a}_2$ thus defines a limit of detection, which can be larger than the photon shot noise. 
It is important to note that this criterion is putting a condition on the variance of $\vec{a}_1$, not on the peak-to-peak amplitude $a_\text{pp}$. Since $\vec{a}_1$ is a sparse signal of $N_\text{ap}$ unit peaks in an otherwise zero vector, $\sqrt{\text{var}(\vec{a}_{1})} =\sqrt{\frac{N_{\text{ap}}}{N_{\text{fr}}}}a_\text{pp}$, and
the criterion for successful signal recovery becomes:

\[
\sigma_{\text{tot}}<\sqrt{\frac{N_{\text{ap}}}{N_{\text{fr}}}} a_\text{pp}.
\] 
Here $a_\text{pp}$ is peak-to-peak amplitude of AP signal. Note that we still assume this to be one, so that $\sigma_i$ denotes the strength of cell noise relative to an action potential.

In a realistic scenario where $N_{\text{ap}}=10$ and $N_{\text{fr}}=1000$, this implies that amplitude of AP signal must even be 10 times higher than the cell noise to recover the signal. 


When $\sigma_\text{tot}$ is distributed among multiple components ($N_\text{c}$ > 2), a higher total amount of $\sigma_\text{tot}$ can be tolerated, if the noise is spread sufficiently evenly across all modes of motion. Still, the signal has to be the largest eigenvector for recovery, and the recovery criterion becomes:

\[
\max(\sigma_{i})<\sqrt{\frac{N_{\text{ap}}}{N_{\text{fr}}}} a_\text{pp}.
\]

Additional insights can be obtained from a  geometrical intuition. The high-dimensional noise can be seen seen as a "blob" in high-dimensional space, where total variance is a sum of variances along every dimension, while the first principal component is the longest axis of that "blob". Both coincide when $N_c=2$, but the variance of largest noise direction is getting smaller as number of dimensions increases and the total noise $\sigma_\text{tot}$ is distributed among them (as a consequence of the triangle inequality). It is difficult to determine the exact lower bound for the largest noise variance, but if variance is distributed highly unevenly evenly among components, it will be on the order of t
he largest component variance, resulting in the recovery criterion above.


\section{Beyond the dominant eigenvector: ideas for an improved algorithm}
Does the signal have to be the largest eigenvector to be recovered? Interestingly, this might not be strictly required, as can be seen in the case $N_c = 2$. 

\begin{figure}[htbp]
\centering\includegraphics[width=\textwidth]{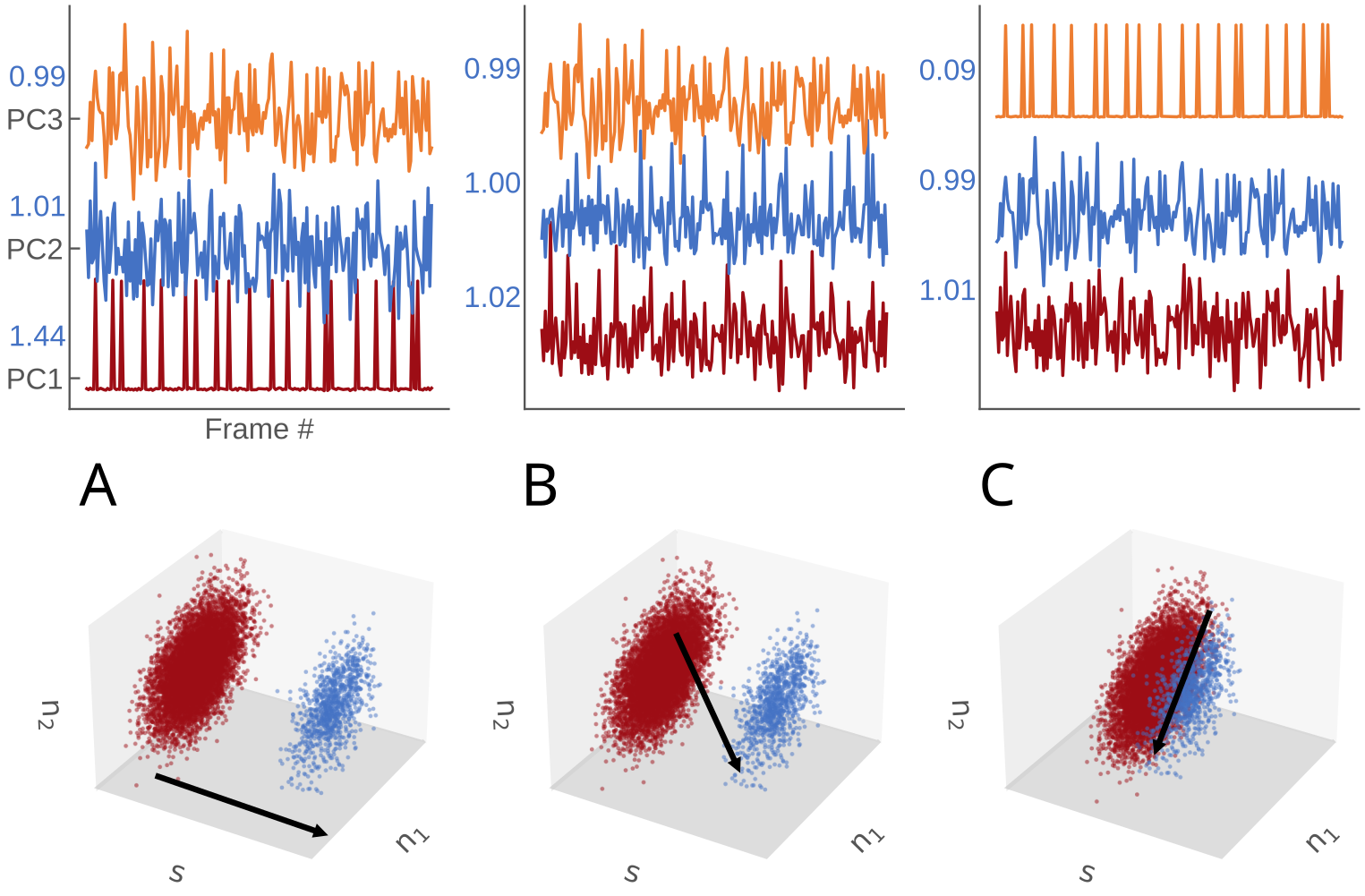}


\caption{Numerical simulation for \( N_{\text{c}}=3 \), demonstrating signal scale variation. (a) strong signal (\(\text{var}(\vec{a}_1) > \text{var}(\vec{a}_{2,3})\)); (b) moderate signal (\(\text{var}(\vec{a}_1) \approx \text{var}(\vec{a}_{2,3})\)); (c) weak signal (\(\text{var}(\vec{a}_1) < \text{var}(\vec{a}_{2,3})\)). The upper row shows the reconstructed principal components, with explained variance for each PC indicated in blue text. The lower row displays amplitudes of the signal (\(\vec{s}=\vec{a}_1\)) and noise (\(\vec{n}_1=\vec{a}_2\), \(\vec{n}_2=\vec{a}_3\)) as a 3D scatter plot, where each dot represents a frame (i.e. an entry of \(\vec{a}_i\)). \\ The black arrow indicates the alignment of the first principal component (PC1). When the signal variance exceeds the noise variance (a), PC1 aligns with the \(\vec{s}\) axis, allowing for perfect signal reconstruction as PC1. When the signal and noise variances are similar (b), PC1 aligns diagonally, distributing the signal across multiple principal components. When the signal variance is much lower (c), PC1 lies within the \(\vec{n}_1\)-\(\vec{n}_2\) plane, restoring the signal as PC3. Data points are color-coded based on AP presence for better visualization.
}

\label{fig:pca_in_3d}
\end{figure}

Here, even if the noise vector has greater variance and thus makes up the first principal component (PC1), the orthogonality of the signal and noise ensures that the second principal component (PC2) will capture the remaining variability, thus perfectly reconstructing the signal. This principle is illustrated in Fig.~\ref{fig:pca_in_3d}. More precisely, the signal can be completely restored when its variance is different from the variance of other components, a principle known as eigengap. A significant eigengap, such as when all noise components are substantially larger than the signal, allows the removal of noise by subtracting $N_{c}-1$ principal components, leading to a perfect restoration of the signal. If shot noise is taken into account, the signal will be corrupted by it. In this case, however, the problem is basically reduced to the one described in the previous section, as all of the cell noise components can be perfectly subtracted.

In practical scenarios involving cells containing numerous scatterers of varying sizes and motion amplitudes, some scatterers have the same variance as the signal, making the recovery much more difficult. Thus, it can be concluded that effective signal detection using a PCA-like approach requires opening the eigengap, i.e. ensuring that cell noise components have significantly different amplitude that AP signal. Counterintuitively, one of the ways to do it is to artificially increase the cell noise components to have variance above the signal variance.

\section{Outlook of other potential approaches}

So far it has been demonstrated that while PCA-like approaches are highly effective when only shot noise is present (or, more generally, in presence of i.i.d. uncorrelated noise), their performance deteriorates for when cell noise is present. Other approaches can potentially be more beneficial under assumption of cell noise dominance. Below we summarize the most promising approaches, detailed analysis of which are beyond the scope of this article.

\paragraph{Independent component analysis (ICA)}
ICA is looking  for a projection of high-dimensional signals which would maximize non-Gaussianity of the recovered signals. There are physics-motivated reasons to assume that the only highly non-Gaussian component of signal is action potential, since all other noises should have approximately Gaussian distribution. Moreover, it should be possible to use variation of ICA which maximizes skewness as a metric of non-Gaussianity instead of usually used kurtosis or negentropy \cite{hyvarinen2000independent}. This version of algorithm \cite{song2016ecoica} is expected to better reflect the nature of the signal, since AP is highly non-symmetric. Another property of the signal which can be exploited is it's sparsity, suggesting usage of ICA variations that specifically optimize for sparsity \cite{chichocki2004beyond}.

\paragraph{Computer vision algorithms}
In contrast to ChiSCAT, which compromises visual interpretability for enhanced spatial resolution, iSCAT provides visually interpretable images. It was observed during experiments that, sometimes, a naked eye can recognize a fringe-like pattern caused by light reflecting off an entire cell. Just as the classical iSCAT algorithm \cite{park2018label} can track individual particles, it should be feasible to track fringes from the cell outer membrane, thereby selecting the action potential (AP) signal using prior knowledge of the spatial patterns associated with it. This tracking can be accomplished using classical computer vision techniques such as the Hough transform \cite{shapiro1996hough} and wavelet transform \cite{wang2012recent}, or by employing modern deep learning methods that are capable of both detecting fringe patterns \cite{reyes2021deep} and reconstructing full wavefronts (unwrapping) \cite{feng2019fringe}. Ultimately, a full refractive index tomography could be performed, so that individual objects in a cell could be tracked. 


\section{Conclusion}

This article provides a theoretical analysis of the criteria for AP detection together with experimental findings about noise properties of different laser sources. We found that use of low-coherence illumination strongly reduces vibrational noise and that a well-optimized setup can indeed operate close to the fundamental photon shot noise limit. However, our theoretical investigation also highlights that the noise from cellular motions is a much more significant challenge for AP detection than shot noise. Theoretical analysis suggests that detection of AP with ChiSCAT can be achieved by opening the eigengap, i.e. modifying the amplitude of the noise to make it significantly different to the amplitude of the signal, which could define a direction for future work.
Alternative strategies, such as ICA and advanced computer vision techniques, also have potential to overcome the limitations associated with cellular noise. These methods, which exploit non-Gaussianity and sparseness of action potentials, or \textit{apriori} information about spatial fringe patterns associated with the cell, may allow reliable AP detection. Future research is needed to explore their integration into ChiSCAT and related interferometric microscopy techniques and experimentally evaluate their performance. If successful, label-free and non-invasive AP detection in living tissues will pave the way for new insights in cellular biology and neuroscience.

\begin{backmatter}
\bmsection{Disclosures}
The authors declare no conflicts of interest.\end{backmatter}

\bibliography{sample}

\begin{thebibliography}{10}
\newcommand{\enquote}[1]{``#1''}

\bibitem{hill49}
D.~K. Hill and R.~D. Keynes, \enquote{Opacity changes in stimulated nerve,} {\protect\JournalTitle{The Journal of Physiology}} \textbf{108}, 278--281 (1949).

\bibitem{cohen68}
L.~B. Cohen, R.~D. Keynes, and B.~Hille, \enquote{Light {Scattering} and {Birefringence} {Changes} during {Nerve} {Activity},} {\protect\JournalTitle{Nature}} \textbf{218}, 438--441 (1968).

\bibitem{iwasa80}
K.~Iwasa, I.~Tasaki, and R.~C. Gibbons, \enquote{Swelling of {Nerve} {Fibers} {Associated} with {Action} {Potentials},} {\protect\JournalTitle{Science}} \textbf{210}, 338--339 (1980). Publisher: American Association for the Advancement of Science.

\bibitem{laporta1990recording}
A.~La~Porta, R.~Stepnoski, F.~Raccuia-Behling, \emph{et~al.}, \enquote{Recording action potential in cultured aplysia neurons using intrinsic optical signals,} in \emph{OSA Annual Meeting,}  (Optica Publishing Group, 1990), p. MPP3.

\bibitem{stepnoski1991noninvasive}
R.~Stepnoski, A.~LaPorta, F.~Raccuia-Behling, \emph{et~al.}, \enquote{Noninvasive detection of changes in membrane potential in cultured neurons by light scattering.} {\protect\JournalTitle{Proceedings of the National Academy of Sciences}} \textbf{88}, 9382--9386 (1991).

\bibitem{macvicar1991imaging}
B.~A. MacVicar and D.~Hochman, \enquote{Imaging of synaptically evoked intrinsic optical signals in hippocampal slices,} {\protect\JournalTitle{Journal of Neuroscience}} \textbf{11}, 1458--1469 (1991).

\bibitem{kim2007mechanical}
G.~Kim, P.~Kosterin, A.~Obaid, and B.~Salzberg, \enquote{A mechanical spike accompanies the action potential in mammalian nerve terminals,} {\protect\JournalTitle{Biophysical journal}} \textbf{92}, 3122--3129 (2007).

\bibitem{laporta2012interferometric}
A.~LaPorta and D.~Kleinfeld, \enquote{Interferometric detection of action potentials,} {\protect\JournalTitle{Cold Spring Harb. Protoc}} \textbf{2012}, 307--311 (2012).

\bibitem{badreddine2016real}
A.~H. Badreddine, T.~Jordan, and I.~J. Bigio, \enquote{Real-time imaging of action potentials in nerves using changes in birefringence,} {\protect\JournalTitle{Biomedical optics express}} \textbf{7}, 1966--1973 (2016).

\bibitem{yang2018imaging}
Y.~Yang, X.-W. Liu, H.~Wang, \emph{et~al.}, \enquote{Imaging action potential in single mammalian neurons by tracking the accompanying sub-nanometer mechanical motion,} {\protect\JournalTitle{ACS nano}} \textbf{12}, 4186--4193 (2018).

\bibitem{ling2018full}
T.~Ling, K.~C. Boyle, G.~Goetz, \emph{et~al.}, \enquote{Full-field interferometric imaging of propagating action potentials,} {\protect\JournalTitle{Light: Science \& Applications}} \textbf{7}, 107 (2018).

\bibitem{ling2020high}
T.~Ling, K.~C. Boyle, V.~Zuckerman, \emph{et~al.}, \enquote{High-speed interferometric imaging reveals dynamics of neuronal deformation during the action potential,} {\protect\JournalTitle{Proceedings of the National Academy of Sciences}} \textbf{117}, 10278--10285 (2020).

\bibitem{mueller2014quantitative}
J.~K. Mueller and W.~J. Tyler, \enquote{A quantitative overview of biophysical forces impinging on neural function,} {\protect\JournalTitle{Physical biology}} \textbf{11}, 051001 (2014).

\bibitem{mosbacher1998voltage}
J.~Mosbacher, M.~Langer, J.~H{\"o}rber, and F.~Sachs, \enquote{Voltage-dependent membrane displacements measured by atomic force microscopy,} {\protect\JournalTitle{The Journal of general physiology}} \textbf{111}, 65--74 (1998).

\bibitem{gonzalez2016solitary}
A.~Gonz{\'a}lez-P{\'e}rez, L.~D. Mosgaard, R.~Budvytyte, \emph{et~al.}, \enquote{Solitary electromechanical pulses in lobster neurons,} {\protect\JournalTitle{Biophysical chemistry}} \textbf{216}, 51--59 (2016).

\bibitem{trelin2024chiscat}
A.~Trelin, S.~Kussauer, P.~Weinbrenner, \emph{et~al.}, \enquote{Chiscat: Unsupervised learning of recurrent cellular micromotion patterns from a chaotic speckle pattern,} {\protect\JournalTitle{Nano Letters}}  (2024).

\bibitem{kukura2009high}
P.~Kukura, H.~Ewers, C.~M{\"u}ller, \emph{et~al.}, \enquote{High-speed nanoscopic tracking of the position and orientation of a single virus,} {\protect\JournalTitle{Nature methods}} \textbf{6}, 923--927 (2009).

\bibitem{perry2018optimality}
A.~Perry, A.~S. Wein, A.~S. Bandeira, and A.~Moitra, \enquote{Optimality and sub-optimality of pca i: Spiked random matrix models,} {\protect\JournalTitle{The Annals of Statistics}} \textbf{46}, 2416--2451 (2018).

\bibitem{ghez1991cerebellum}
C.~Ghez, \enquote{{The Cerebellum},} {\protect\JournalTitle{{Principles of Neural Science}}} pp. 626--646 (1991).

\bibitem{hyvarinen2000independent}
A.~Hyv{\"a}rinen and E.~Oja, \enquote{Independent component analysis: algorithms and applications,} {\protect\JournalTitle{Neural networks}} \textbf{13}, 411--430 (2000).

\bibitem{song2016ecoica}
L.~Song and H.~Lu, \enquote{Ecoica: Skewness-based ica via eigenvectors of cumulant operator,} in \emph{Asian Conference on Machine Learning,}  (PMLR, 2016), pp. 445--460.

\bibitem{chichocki2004beyond}
A.~Chichocki, Y.~Li, P.~Georgiev, and S.-i. Amari, \enquote{Beyond ica: Robust sparse signal representations,} in \emph{2004 IEEE International Symposium on Circuits and Systems (ISCAS),}  vol.~5 (IEEE, 2004), pp. V--V.

\bibitem{park2018label}
J.-S. Park, I.-B. Lee, H.-M. Moon, \emph{et~al.}, \enquote{Label-free and live cell imaging by interferometric scattering microscopy,} {\protect\JournalTitle{Chemical Science}} \textbf{9}, 2690--2697 (2018).

\bibitem{shapiro1996hough}
V.~A. Shapiro, \enquote{On the hough transform of multi-level pictures,} {\protect\JournalTitle{Pattern Recognition}} \textbf{29}, 589--602 (1996).

\bibitem{wang2012recent}
Z.~Wang, J.~Ma, and M.~Vo, \enquote{Recent progress in two-dimensional continuous wavelet transform technique for fringe pattern analysis,} {\protect\JournalTitle{Optics and Lasers in Engineering}} \textbf{50}, 1052--1058 (2012).

\bibitem{reyes2021deep}
A.~Reyes-Figueroa, V.~H. Flores, and M.~Rivera, \enquote{Deep neural network for fringe pattern filtering and normalization,} {\protect\JournalTitle{Applied Optics}} \textbf{60}, 2022--2036 (2021).

\bibitem{feng2019fringe}
S.~Feng, Q.~Chen, G.~Gu, \emph{et~al.}, \enquote{Fringe pattern analysis using deep learning,} {\protect\JournalTitle{Advanced photonics}} \textbf{1}, 025001--025001 (2019).

\end{thebibliography}






\end{document}